\documentclass[twocolumn,showpacs,preprintnumbers,amsmath,amssymb,prb]{revtex4}

\usepackage{graphicx}
\usepackage{dcolumn}
\usepackage{bm}
\usepackage{epsfig}
\newcommand{\be}{\begin{eqnarray}}
\newcommand{\ee}{\end{eqnarray}}

\begin{document}

\title{Phase Transitions and Pairing Signature in Strongly Attractive Fermi Atomic Gases}

\author{ X.W. Guan$^{\dagger \ddagger}$, M.T. Batchelor$^{\dagger  \ddagger}$,
  C. Lee$^{\star}$ and M. Bortz$^{\dagger}$}
\affiliation{${\dagger}$ Department of Theoretical Physics, Research
  School of Physical Sciences and Engineering, 
Australian National University, Canberra ACT 0200,  Australia}

\affiliation{${\ddagger}$ Mathematical Sciences Institute, 
Australian National University, Canberra ACT 0200,  Australia}

\affiliation{$\star$ Nonlinear Physics Centre and ARC Centre of Excellence
for Quantum-Atom Optics, Research School of Physical Sciences and Engineering\\
Australian National University, Canberra ACT 0200,
Australia}

\date{\today}

\begin{abstract}

We investigate pairing and quantum phase transitions in the
one-dimensional two-component Fermi atomic gas in an external
field. The phase diagram, critical fields, magnetization and local
pairing correlation are obtained analytically via the exact
thermodynamic Bethe ansatz solution. At zero temperature, bound pairs of
fermions with opposite spin states form a singlet ground state when
the external field $H < H_{c1}$. A completely ferromagnetic phase
without pairing occurs when the external field $H > H_{c2}$.  In the
region $H_{c1} < H < H_{c2}$ we observe a mixed phase of matter in
which paired and unpaired atoms coexist. The phase diagram is
reminiscent of that of type II superconductors.  For temperatures
below the degenerate temperature and in the absence of an external field,
the bound pairs of fermions form hard-core bosons obeying generalized
exclusion statistics.

\end{abstract}



\maketitle

\section{Introduction}

Recent achievements in manipulating quantum gases of ultracold
atoms have opened up exciting possibilities for the experimental
study of many-body quantum effects in low-dimensional systems \cite{Lewenstein,Grimm,Demler}.
Experimental observation of superfluidity and phase separation in
imbalanced Fermi atomic gases \cite{MIT_imbalanced-Fermi-gas,Rice_imbalanced-Fermi-gas} have
stimulated great interest in exploring exotic quantum phases of matter
with two mismatched Fermi surfaces.
The pairing of fermionic atoms with mismatched Fermi surfaces
may lead to a breached pairing phase \cite{Sama} and a nonzero
momentum pairing phase of Fulde-Ferrell-Larkin-Ovchinnikov (FFLO) states \cite{FFLO}.
In general the nature of pairing and superfluidity in strongly interacting systems 
is both subtle and intriguing \cite{Schunck}.

Pairing is well known to be a momentum space phenomenon, 
in which two fermions with opposite spin states form a bound pair  
which behaves like a boson. 
The bound pairs form a superfluid, while the unpaired fermions
remain as a separated gas phase in momentum space. 
Such superfluid states with gapless excitations in ultracold atomic gases 
provide an exciting insight into the superfluid regime in quantum many-body physics.  
Fermi gases of ultracold atoms with population imbalance have been
predicted to exhibit a quantum phase transition between the normal and
superfluid states \cite{MF-FFLO,Wilczek2,Dukelsky,Cazalilla}.
Mismatched Fermi surfaces can appear in different quantum systems,
such as type II superconductors in an external magnetic field
\cite{Super-II}, a mixture of two species of fermionic atoms with
different densities or masses \cite{MF-FFLO,Wilczek2} and 
charge neutral quark matter \cite{ARW,RZHH}.

These exotic phases have attracted newfound interest
in the one-dimensional (1D) integrable two-component Fermi gas
\cite{Yang,Gaudin,Takahashi} which was used to study BCS-BEC crossover
\cite{BCS-BEC,Wadati} and quantum phase separation in a trapping
potential \cite{Orso,Hu}.
The 1D Fermi gases can be experimentally realized by applying
strongly transverse confinement to the Fermi atomic clouds
\cite{Fermi-1D1}.
In the 1D  interacting Fermi gas, the Fermi surface is reduced to the Fermi
points. The lowest excitation destroys a bound pair close to the
Fermi surface. 
Charge and spin propagate with different velocities due to the
pair-wise interaction.
The external magnetic field triggers energy level crossing such that  the
Fermi surfaces of paired fermions and unpaired fermions vary smoothly
with respect to the external field. 
As we shall see in this paper, the presence of the external
field at zero temperature has an important bearing on the nature of 
quantum phase transitions in 1D interacting fermions.

In general the exact Bethe ansatz (BA) solution of any model provides reliable
physics beyond mean field theory \cite{Batchelor}.
The thermodynamic Bethe ansatz (TBA)
\cite{Y-Y,Kondo,Takahashi,Schlot,Hubbardbook,BGOS} provides a
way to obtain the ground state signature and finite temperature
behaviour of integrable 1D quantum many-body systems.
At zero temperature, the TBA equations naturally reduce to
dressed energy equations in which the external field is explicitly involved. 
Thus the band fillings are subsequently varied with respect to the external
field.
This gives an elegant way to analyze quantum phase transitions in the
presence of an external field by means of the dressed energy formalism.
Our aim here is to obtain new exact results from this
formalism for characteristics of pairing phases and quantum phase
transitions in the 1D two-component strongly attractive Fermi gas of
cold atoms.
We present a systematic way to obtain the critical
fields and magnetic properties at zero temperature for strongly interacting fermions.
 We find that the bound pairs of
fermions with opposite spin states form a singlet ground state when
the external field $H < H_{c1}$.
A completely ferromagnetic phase
without pairing occurs when the external field $H > H_{c2}$.
In the region $H_{c1} < H < H_{c2}$ we observe a mixed phase of matter in
which paired and unpaired atoms coexist. 
However, in the absence of the external magnetic field, we show that the bound
pairs of fermions behave like hard-core diatoms obeying nonmutual
generalized exclusion statistics (GES) at temperatures much less than the 
binding energy. 

This paper is set out as follows. In section \ref{model}, we present the BA solution of the 1D
two-component interacting Fermi gas. The ground state properties are
also analysed. In section \ref{TBA}, we introduce the TBA in order to set 
up the dressed energy formalism. The
quantum phase transitions and magnetic properties are studied by
means of the dressed energy formalism in section \ref{QP}. We discuss
the distribution profiles and the thermodynamics of the 1D strongly
attractive Fermi gas of atoms at low temperatures in section \ref{GES}, 
along with the connection to GES.   
Section \ref{conclusion} is devoted to concluding remarks.


\section{The model}
\label{model}
The model we consider has interacting atoms in two
hyperfine levels $|1\rangle$ and $|2\rangle$, which are coherently
coupled with laser or Rabi frequency fields.
Under strong transverse confinement, the system is effectively
described along the axial direction by the 1D Hamiltonian 
${\cal H} = {\cal H}_{0} +{\cal H}_{int} +{\cal H}_{c}$.
The first term 
\begin{equation}
{\cal H}_{0} = \sum_{j=1}^{2} \int \psi _{j}^{\dagger}(x) \left
(-\frac{\hbar^{2}}{2m}\frac{d^{2}}{dx^{2}} + V(x) \right ) \psi_{j}^{}(x) dx
\end{equation}
contains the kinetic energy and the trapping potential $V(x)$.
The second term
\begin{equation}
{\cal H}_{int} = g_{1D} \int \psi _{1}^{\dagger}(x) \psi _{2}^{\dagger}(x) \psi _{2}^{}(x) \psi_{1}^{}(x) dx
\end{equation}
describes $s$-wave interaction and 
\begin{equation}
{\cal H}_{c} = \frac12 {\Omega}
\int \left (\psi _{2}^{\dagger}(x) \psi _{1}^{}(x) + \psi_{1}^{\dagger}(x) \psi _{2}^{}(x) \right ) dx
\end{equation}
is the coupling term.
Here $\psi^{\dagger}_{1}(x)$ and $\psi^{\dagger} _{2}(x)$ are the
atomic field creation operators, $g_{1D}$ is the 1D interaction
strength and $\Omega$ is the Rabi frequency of coupling fields.

Defining
$\psi_{1}=(\phi_{\downarrow}+\phi_{\uparrow})/\sqrt{2}$ and
$\psi_{2}=(\phi_{\downarrow}-\phi_{\uparrow})/\sqrt{2}$, the
Hamiltonian becomes
\begin{eqnarray}
{\cal H} &=& \sum _{j=\downarrow,\uparrow} \int \phi _{j}^{\dagger}(x) \left
(-\frac{\hbar^{2}}{2m}\frac{d^{2}}{dx^{2}} + V(x) \right ) \phi
_{j}^{}(x) dx\nonumber\\
&& +  g_{1D} \int \phi _{\downarrow}^{\dagger}(x) \phi
_{\uparrow}^{\dagger}(x) \phi _{\uparrow}^{}(x)
\phi _{\downarrow}^{}(x) dx \nonumber\\
&& - \frac12{\Omega} \int \left (\phi _{\uparrow}^{\dagger}(x) \phi
_{\uparrow}^{}(x) - \phi _{\downarrow}^{\dagger}(x) \phi
_{\downarrow}^{}(x) \right ) dx. \label{Ham-1}
\end{eqnarray}
The new field operators $\phi_{\downarrow}$ and $\phi_{\uparrow}$
describe the atoms in the states $\left | \downarrow \right \rangle =
(|1\rangle + |2\rangle)/\sqrt{2}$ and $\left | \uparrow \right
\rangle=(|1\rangle-|2\rangle)/\sqrt{2}$. This
Hamiltonian also describes the 1D $\delta$-interacting
spin-$\frac{1}{2}$ Fermi gas with an external magnetic field $H=\Omega$.
Here we consider the homogeneous case $V(x)=0$
with periodic boundary conditions for a line of length $L$ \cite{Yang,Gaudin}.
Unless specifically indicated, we use units of $\hbar =2m=1$.
We define $c =mg_{1D}/\hbar ^2$ and a dimensionless interaction strength
$\gamma=c/n$ for the physical analysis, with linear density $n= {N}/{L}$, 
where $N$ is the number of fermions.
The inter-component interaction can be tuned from strongly
attractive ($g_{\rm 1D}\rightarrow -\infty$) to strongly repulsive
($g_{\rm 1D} \rightarrow +\infty$) via Feshbach resonances.

The model was solved by nested BA \cite{Yang,Gaudin} for the energy eigenspectrum 
\begin{equation}
E=\frac{\hbar ^2}{2m}\sum_{j=1}^N k_j^2
\end{equation}
in terms of the $N$ BA wave numbers $\left\{k_i\right\}$,  which are the quasimomenta of the fermions.
They satisfy the BA equations \cite{Yang,Gaudin}
\begin{eqnarray}
& &\exp(\mathrm{i}k_jL)=\prod^M_{\ell = 1} 
\frac{k_j-\Lambda_\ell+\mathrm{i}\, c/2}{k_j-\Lambda_\ell-\mathrm{i}\, c/2},\nonumber\\
& &\prod^N_{\ell = 1}\frac{\Lambda_{\alpha}-k_{\ell}+\mathrm{i}\, c/2}{\Lambda_{\alpha}-k_{\ell}-\mathrm{i}\, c/2}
 = - {\prod^M_{ \beta = 1} }
\frac{\Lambda_{\alpha}-\Lambda_{\beta} +\mathrm{i}\, c}{\Lambda_{\alpha}-\Lambda_{\beta} -\mathrm{i}\, c} .
\label{BE}
\end{eqnarray}
Here $j = 1,\ldots, N$ and $\alpha = 1,\ldots, M$, with $M$ the number of spin-down fermions.
The additional parameters $\left\{\Lambda_{\alpha}\right\}$ are the rapidities
for the internal spin degrees of freedom.

\begin{figure}[ht]
{{\includegraphics [width=0.98\linewidth]{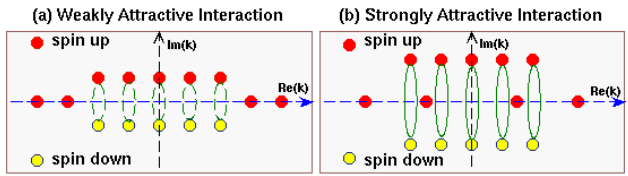}}}
\caption{Schematic Bethe ansatz configuration of quasimomenta $k$ in the complex plane.
(a) For weakly attractive interaction, the quasimomenta of unpaired fermions
sit in the outer wings of the distribution. (b) For strongly attractive interaction,
they  can penetrate into the central region.}
\label{fig:Cartoon}
\end{figure}

The distribution of the quasimomenta in the complex plane was studied recently \cite{BBGO}.
For weakly attractive interaction, the system describes weakly bound
Cooper pairs where the quasimomenta are 
distributed in a BCS-like manner \cite{BBGO} (Figure \ref{fig:Cartoon}(a)).
In this limit, the ground state energy per unit length is given by 
\begin{equation}
E \approx
\frac{\hbar^2n^3}{2m}\left(\frac{\gamma}{2}(1-P^2)+\frac{\pi^2}{12}+\frac{\pi^2}{4}P^2\right),
\end{equation}
with the  polarization $P=(N-2M)/N$. 
The bound state has a small
binding energy $\epsilon_{\rm b}={\hbar^2n|\gamma|}/{m}$ and is
therefore unstable against  thermal fluctuations.
For strongly attractive interaction,  the bound pairs form hard-core bosons
(Figure \ref{fig:Cartoon}(b)). 
The energy per unit length derived directly from Eqs (\ref{BE}) is \cite{BBGO}
\begin{eqnarray}
E &\approx  & \frac{\hbar^2n^3}{2m} \left\{-\frac{(1-P)\gamma^2}{4} +
\frac{P^3\pi^2}{3} \left(1+\frac{4(1-P)}{|\gamma|}\right)\right.\nonumber\\
& & \left. + \frac{\pi^2(1-P)^3}{48}
\left(1+\frac{(1-P)}{|\gamma|}+\frac{4P}{|\gamma|}\right)\right\}
\label{E-P}
\end{eqnarray}
with binding energy $\epsilon_{\rm b} = {\hbar^2n^2\gamma^2}/ {(4m)}$.
Generally, the total momentum for bound pairs and that for unpaired fermions are both zero.
Thus the BA roots for the model with population imbalance do not
show sufficient evidence  for the
existence of a FFLO state which might exist in the asymmetric BCS
pairing models \cite{MF-FFLO,Dukelsky}. 
In FFLO states the unpaired
fermions have an asymmetric distribution at the Fermi surface 
resulting in a net total momentum for bound pairs \cite{FFLO}.

\section{Thermodynamic Bethe Ansatz}
\label{TBA}
The thermodynamic Bethe ansatz (TBA) has been well established in quantum
integrable systems \cite{Y-Y,Kondo,Takahashi,Schlot,Hubbardbook,BGOS}. 
For the sake of completeness, we sketch the main idea and results of the TBA
for the fermion model in this section.  
At zero temperature, all quasimomenta $k_i$ of $N$ atoms form two-body bound states,
i.e., $k_j=\Lambda_j'\pm   \mathrm{i} \frac{1}{2} c$, accompanied by the real spin parameter
 $\Lambda'_j$. 
 Here $j=1,\ldots,M$.  However, at finite temperature,
 spin quasimomenta form complex strings
 $\Lambda_{\alpha,j}^n=\Lambda^n_{\alpha}+\mathrm{i} \frac12 (n+1-2j)c$ with
 $j=1,\ldots ,n$ \cite{Takahashi}. 
 Here the number of strings $\alpha=1,\ldots, N_n$. 
$\Lambda^n_{\alpha}$ is the position of the center
 for the length-$n$ string on the real axis. The number of $n$-strings
 $N_n$ satisfies the relation $M=M'+\sum_nnN_n$. 
 There are $M'$ real $\Lambda'_j$ and there are $N-2M'$ real $k_i$ for unpaired
 fermions.  In the presence of a magnetic field, the ground state
 consists of two Fermi seas: one contains bound pairs, another
 contains unpaired fermions. 
 
In the thermodynamic limit, i.e., $N,L\to
 \infty$ with $N/L$ finite, it is assumed that the distributions of
 Bethe roots is sufficiently dense along the real axis.  
 After introducing
 the root distribution functions $\sigma(k)$, $\rho(k)$ and
 $\xi_n(k)$ for paired fermions, unpaired fermions and the spin
 $n$-string as well as their hole densities $\sigma^h(k)$,
 $\rho^h(k)$ and $\xi^h_n(k)$, the BA equations (\ref{BE})
 can be transformed into the form \cite{Takahashi}
\begin{eqnarray}
\sigma(k)+\sigma^h(k)&=&\frac{1}{\pi}-a_2*\sigma(k)-a_1*\rho(k),\nonumber\\
\rho(k)+\rho^h(k)&=&\frac{1}{2\pi}-a_1*\sigma(k)-\sum_{n=1}^{\infty}a_n*\xi_n(k),\nonumber\\
\xi_n(\Lambda)+\xi_n^h(\Lambda)&=&a_n*\rho(\Lambda)-\sum_{n=1}^{\infty}T_{nm}*\xi_n(\Lambda).\label{BE-string}
\end{eqnarray}
Here $*$ denotes the convolution integral
$(f*g)(\lambda) = \int_{-\infty}^\infty f(\lambda-\lambda')
g(\lambda') d\lambda'$ and 
\begin{equation}
a_m(\lambda)=\frac{1}{2\pi}\frac{m|c|}{(mc/2)^2+\lambda ^2}.\label{a-n}
\end{equation}
The function $T_{nm}(\lambda)$ can be found in Takahashi's book
\cite{Takahashi}.  

The equilibrium states at finite temperature $T$
are described by the equilibrium quasiparticle and hole densities.
The partition function $Z=tr(\mathrm{e}^{-\cal{H}/T})$ is defined as
\begin{eqnarray}
Z&=&\sum_{\sigma,\sigma^h,\rho,\rho^h,\xi_n,\xi_n^h}W
\mathrm{e}^{-E(\sigma,\sigma^h,\rho,\rho^h,\xi_n,\xi_n^h)/{T}},
\end{eqnarray}
where the densities satisfy the BA  equations (\ref{BE-string}), and
$W := W(\sigma,\sigma^h,\rho,\rho^h,\xi_n,\xi_n^h)$ is the number
of states corresponding to the given densities.  By introducing the
combinatorial entropy $S=\ln W$ the grand partition function can be
presented as $Z=\mathrm{e}^{-G/T}$, where
the Gibbs free energy $G=E-\mu N-HM^z-TS$. Here $\mu $ is the chemical
potential. The entropy and Gibbs free
energy are given in terms of the BA root distribution functions of
particles and holes for bound pairs and unpaired fermions as well as spin
degrees of freedom.

The energy per unit length is defined by
\begin{eqnarray}
E= \int_{-\infty}^{\infty}
\left(k^2\rho(k)+2(k^2-\frac{c^2}{4})\sigma(k)\right)dk-M^zH.
\end{eqnarray}
Here $H$ is the external magnetic field and $M^z=(N-2M)/(2L)$ denotes
the atomic magnetic momentum per unit length (where the Bohr magneton
$\mu_B$ and the Lande factor are absorbed into the magnetic field
$H$).  

The entropy per unit length is given by \cite{Takahashi}
\begin{eqnarray}
S&=&\int_{-\infty}^\infty\left((\sigma(k)+\sigma^h(k))\ln(\sigma(k)+\sigma^h(k))\right.\nonumber\\
& &\left.-\sigma(k)\ln\sigma(k)-\sigma^h(k)\ln\sigma^h(k)\right)dk\nonumber\\
&
&+\int_{-\infty}^\infty\left((\rho(k)+\rho^h(k))\ln(\rho(k)+\rho^h(k))\right.\nonumber\\
& &\left.-\rho(k)\ln\rho(k)-\rho^h(k)\ln\rho^h(k)\right)dk\nonumber\\
& &+\sum_{n=1}^{\infty}\int_{-\infty}^{\infty}
\left((\xi_n(\lambda)+\xi_n^h(\lambda))\ln(\xi_n(\lambda)+\xi_n^h(\lambda))\right.\nonumber\\
& &\left.-\xi_n(\lambda)\ln\xi_n(\lambda)-\xi^h_n(\lambda)\ln\xi_n^h(\lambda)\right) d\lambda.
\end{eqnarray}
The equilibrium states are determined by the minimization condition of
the Gibbs free energy, which gives rise to a set of coupled nonlinear
integral equations -- the TBA equations \cite{Takahashi}.
In terms of the dressed energies $\epsilon^{\rm b}(k) := T\ln( \sigma(k)/\sigma^h(k) )$ and
$\epsilon^{\rm u}(k) := T\ln( \rho(k)/\rho ^h(k) )$ for paired and unpaired fermions these are 
\begin{eqnarray}
\epsilon^{\rm
  b}(k)&=&2(k^2-\mu-\frac14{c^2})+Ta_2*\ln(1+\mathrm{e}^{-\epsilon^{\rm b}(k)/T} )
  \nonumber\\
& &+ \, Ta_1*\ln(1+\mathrm{e}^{-\epsilon^{\rm u}(k)/{T}})\nonumber\\
\epsilon^{\rm
  u}(k)&=&k^2-\mu-\frac12{H}+Ta_1*\ln(1+\mathrm{e}^{-\epsilon^{\rm b}(k)/{T}})\nonumber\\
& &-T\sum_{n=1}^{\infty}a_n*\ln(1+\eta_n^{-1}(k))\nonumber\\
\ln
  \eta_n(\lambda)&=&\frac{nH}{T}+a_n*\ln(1+\mathrm{e}^{-\epsilon^{\rm u}(\lambda)/{T}})\nonumber
\\&&+\sum_{n=1}^{\infty}T_{mn}*\ln(1+\eta^{-1}_n(\lambda)).\label{TBA-Full}
\end{eqnarray}
The function $\eta_n(\lambda) := \xi(\lambda)/\xi ^h(\lambda ) $ is the ratio of the
string densities.  
The Gibbs free energy per unit length is given by
\begin{eqnarray}
G&=&\frac{T}{\pi}\int_{-\infty}^{\infty}dk\ln(1+\mathrm{e}^{-\epsilon^{\rm
      b}(k)/{T}})\nonumber\\
      &&+\,\frac{T}{2\pi}\int_{-\infty}^{\infty}dk \ln(1+\mathrm{e}^{-\epsilon^{\rm u}(k)/{T}}).\label{pressure}
\end{eqnarray}
The TBA equations
provide a clear picture of band fillings with respect to the field
$H$ and the chemical potential $\mu$ at arbitrary temperatures. 
However, it
is a challenging problem to obtain analytic results for the thermodynamics at
low temperatures from the TBA (\ref{TBA-Full}).

We focus on quantum phase transitions in the 1D strongly attractive
Fermi gas at $T=0$ by analyzing the dressed energy equations
\begin{eqnarray}
\epsilon^{\rm b}(\Lambda)&=&2\left(\Lambda^2-\mu
-\frac{c^2}{4}\right)-\int_{-B}^{B}a_2(\Lambda-\Lambda'){\epsilon^{\rm
    b}}(\Lambda')d\Lambda' \nonumber\\
& &-\int_{-Q}^{Q}a_1(\Lambda-k){\epsilon^{\rm u}}(k)d k,\nonumber\\
\epsilon^{\rm u}(k)&=&\left(k^2-\mu
-\frac{H}{2}\right)-\int_{-B}^{B}a_1(k-\Lambda){\epsilon^{\rm  b}}(\Lambda)d\Lambda
\label{TBA-F}
\end{eqnarray}
which are obtained from the TBA equations (\ref{TBA-Full}) in  the limit $T\to 0$.
The dressed energy $\epsilon^{\rm
b}(\Lambda)\le 0$ ($\epsilon^{\rm u}(k)<0$) for $|\Lambda|\le B$
($|k|\le Q$) correspond to the occupied states. The positive part of
$\epsilon^{\rm b}$ ($\epsilon^{\rm u}$) corresponds to the unoccupied states.
The integration boundaries $B$ and $Q$ characterize the Fermi surfaces for
bound pairs and unpaired fermions, respectively.
The Gibbs free energy per unit length at zero temperature is given by
\begin{equation}
G(\mu,H)=\frac{1}{\pi}\int_{-B}^B{\epsilon^{\rm b}}(\Lambda)d\Lambda+
\frac{1}{2\pi}\int_{-Q}^Q{\epsilon^{\rm u}}(k)dk.
\end{equation}
The magnetization $M^z=nP/2$ per unit length is determined by $H$,
$g_{1D}$ and $n$ through the relations
\begin{eqnarray}
-\partial G(\mu,H)/\partial \mu =n\,,\,\,-\partial
G(\mu,H)/\partial H =M^z\label{deffield}.
\end{eqnarray}

\section{Quantum Phase Transitions}
\label{QP}

The ground state is antiferromagnetic, i.e., the number of the fermionic
atoms with up-spin states and  the number of the fermionic
atoms with down-spin states are equal. In this case the integral limit
for the unpaired Fermi sea $Q=0$ and $\rho(k)=0$. For strong coupling,
i.e. $L|c| \gg 1$ the dressed energy equations (\ref{TBA-F})  reduce to
the form
\begin{eqnarray}
\epsilon^{\rm b}(\Lambda)\approx 2\left(\Lambda^2-\mu
-\frac{c^2}{4}\right)-\frac{1}{2\pi}\int_{-B}^{B}\frac{2|c|\epsilon^{\rm
    b}(\Lambda')d\Lambda'}{c^2+k^2}. \label{TBA-g}
\end{eqnarray}
For convenience of notation we denote 
\begin{eqnarray}
p^{\rm b}&=&-\frac{1}{\pi}\int_{-B}^B{\epsilon^{\rm b}}(\Lambda)d\Lambda,\nonumber\\
p^{\rm u}&=&-\frac{1}{2\pi}\int_{-B}^B{\epsilon^{\rm u}}(\Lambda)d\Lambda, \label{P-b}
\end{eqnarray}
as the pressure for bound pairs and unpaired fermions. 
Substituting equation (\ref{TBA-g}) into
  $p^{\rm b}$, we have 
\begin{equation}
\pi p^{\rm b}\left(1+\frac{2B}{\pi|c|} \right)\approx 4B\left( \mu -\frac{1}{3}B^2+\frac{c^2}{4}\right).\label{TBA-g1}
\end{equation}
Furthermore, from the Fermi points $\epsilon^{\rm b}(\pm B)=0$, we have 
\begin{equation}
B^2\approx \mu+\frac{c^2}{4}-\frac{p^{\rm b}}{2|c|}.\label{TBA-g2}
\end{equation}
{}From the relation (\ref{deffield}) together with the equations
(\ref{TBA-g1}) and (\ref{TBA-g2}), we obtain the pressure and the
ground state energy per unit length as
\begin{eqnarray}
p^{\rm b}& \approx
&\frac{\hbar^2}{2m} \frac{\pi^2 n^3}{24} \left(1+\frac{3}{2|\gamma|}\right),\nonumber\\
E_0& \approx
&\frac{\hbar^2n^3}{2m} \left( -\frac{\gamma^2}{4}+ \frac{1}{48}\pi^2\left(1+\frac{1}{|\gamma|}\right)\right).\label{e0}
\end{eqnarray} 

For the strongly attractive 1D Fermi gas, the low-energy excitations
split into collective excitations carrying charge and collective
excitations carrying spin. 
This leads to the phenomenon of spin-charge separation. 
The spin excitation is gapped with a divergent spin velocity
\begin{equation}
v_s=\frac{n|\gamma|}{\sqrt{2}}\left(1+\frac{2}{|\gamma|}\right).  
\end{equation}
Therefore the spin sector cannot be described by a conformal field theory.
However, the charge sector is still critical  \cite{Yang-K} with central charge
$C=1$ and the charge velocity
\begin{equation}
v_c=\frac{v_{\rm F}}{4}\left(1+\frac{1}{|\gamma|}\right)\label{vc}
\end{equation}
for the fully paired ground state, where the
bound pairs behave like hard-core bosons. 
In the above equation $v_{\rm F}=\hbar n\pi/4m$. The bound pairs can be
broken by a strong enough external field or thermal fluctuations. In the
strong interaction limit, it was demonstrated in a recent experiment
that the nature of the pairing is likely to be molecular and
the mismatched Fermi surfaces do not prevent pairing but indeed quench
the superfluidity \cite{Schunck}. The state with polarization can
be viewed  as an ideal mixture of bosonic pairs and fermionic
quasiparticles.

With polarization $0<P<1$, from the equations (\ref{TBA-F}) we obtain 
\begin{eqnarray}
 p^{\rm b} &\approx &-
 \frac{4B}{\pi}\left(\frac{B^2}{3}-\mu-\frac{c^2}{4}+\frac{p^{\rm
 b}}{2|c|} +\frac{2p^{\rm u}}{|c|}\right),\nonumber\\
p^{\rm u} &\approx &-
 \frac{Q}{\pi}\left(\frac{Q^2}{3}-\mu-\frac{H}{2}+\frac{2p^{\rm b}}{|c|}\right),\label{polarize-1}
\end{eqnarray}
{}From the Fermi points $\epsilon^{\rm b}(B)=0$ and $\epsilon^{\rm u}(Q)=0$
we have 
\begin{eqnarray}
2(B^2-\mu-\frac{c^2}{4})+\frac{p^{\rm b}}{|c|}+\frac{4p^{\rm
      u}}{|c|} &\approx& 0,\nonumber\\
Q^2-\mu-\frac{H}{2}+\frac{2p^{\rm b}}{|c|} &\approx& 0.\label{polarize-2}
\end{eqnarray}
It follows that  
\begin{eqnarray}
 p^{\rm b} &\approx
 &\frac{8}{3\pi}\left(\mu+\frac{c^2}{4}-\frac{p^{\rm
 b}}{2|c|}-\frac{2p^{\rm u}}{|c|} \right)^{\frac{3}{2}},\nonumber\\
 p^{\rm u} &\approx
 &\frac{2}{3\pi}\left(\mu+\frac{H}{2}-\frac{p^{\rm
 b}}{2|c|}\right)^{\frac{3}{2}}.
\end{eqnarray}
With the help of the relation (\ref{deffield}) and by lengthy iteration we
find the effective chemical potentials for pairs $\mu^{\rm b}=\mu+\epsilon_{\rm b}/2 $ 
and unpaired fermions $\mu^{\rm u}=\mu+H/2$ are given by
\begin{equation}
\mu^{\rm b} \approx  \frac{\hbar^2n^2\pi^2}{2m}\left\{\frac{(1-P)^2}{16}\left(1+\frac{4(1-P)}{3|\gamma|}+\frac{4P}{|\gamma|}\right)
+\frac{4P^3}{3|\gamma|}\right\} \label{Mu-b}
\end{equation}
and
\begin{equation}
\mu^{\rm u} \approx \frac{\hbar^2n^2\pi^2}{2m}\left\{P^2\left(1+\frac{4(1-P)}{|\gamma|}\right)+
\frac{(1-P)^3}{12|\gamma|}\right\}.\label{Mu-u}
\end{equation}
These results can give explicit chemical potentials for the two different
species: 
\begin{equation}
\mu_{\uparrow}=\mu+H/2, \,\,\, \mu_{\downarrow}=\mu-H/2.
\end{equation}
In addition we have the total chemical potential $\mu=\partial
E/\partial n-HP/2$. Here the energy per unit length with polarization
(\ref{E-P}) follows from the relation $E =n\mu
-G(\mu,H)+{nHP}/{2}$. Indeed, the energy obtained from the TBA
formalism is in agreement with the result (\ref{E-P}) derived from the
BA \cite{BBGO}.
The integration boundaries 
\begin{eqnarray}
B &\approx& \frac{n\pi(1-P)}{4}
\left(1+\frac{(1-P)}{2|\gamma|} +\frac{2P}{|\gamma|}\right), \nonumber\\
Q &\approx& n\pi P\left( 1 + \frac{2(1-P)}{|\gamma|}\right)
\end{eqnarray}
are the largest quasimomentum for bound pairs and unpaired fermions.

Analysis of the dressed energy equations (\ref{TBA-F}) shows that the
fully paired ground state with $M^{z} = 0$ is stable when the field
$H<H_{c1}$, where
\begin{equation}
H_{c1} \approx \frac{\hbar^2n^2}{2m} \left( \frac{\gamma^2}{2} -
\frac{\pi^2}{8} \right).
\label{Hc1}
\end{equation}
This critical field makes the excitation gapless. If the external
field $H>H_{c1}$, the {\it pairing gap}, defined by
$\Delta=(H_{c1}-H)/2$, is completely diminished by the external field.
Slightly above the
critical point $H_{c1}$, the system has a linear field-dependent magnetization 
\begin{equation}
M^z\approx \frac{2(H-H_{c1})}{n\pi^2}\left(1+\frac{2}{|\gamma|}\right)
\end{equation}
with a finite susceptibility 
\begin{equation}
\chi \approx \frac{2}{n\pi^2}\left(1+\frac{2}{|\gamma|}\right).
\end{equation}
This behaviour differs from the Pokrovsky-Talapov-type phase transition
occurring in a gapped spin liquid.
We note that this smooth phase transition in the attractive Fermi gas
is reminiscent of the transition from the Meissner phase to the mixed
phase in type II superconductors \cite{Super-II}.

On the other hand, if the external field $H>H_{c2}$, where
\begin{equation}
H_{c2} \approx  \frac{\hbar^2n^2}{2m} \left( \frac{\gamma^2}{2} +
2\pi^2\left(1-\frac{4}{3|\gamma|}\right) \right)
\label{Hc2}
\end{equation}
all bound pairs are broken and the ground state becomes a normal
ferromagnetic state of fully polarized atoms.
Slightly below $H_{c2}$ the phase transition is determined by the linear field-dependent relation
\begin{equation}
M^z \approx \frac12{n}\left(1-\frac{(H_{c2}-H)}{4n^2\pi^2}\left(1+\frac{10}{3|\gamma|}\right)\right)
\end{equation}
with a finite susceptibility 
\begin{equation}
\chi \approx \frac{1}{8n\pi^2}\left(1+\frac{10}{3|\gamma|}\right).
\end{equation}

A mixed phase occurs in the region $H_{c1}< H< H_{c2}$, with coexistence of spin singlet bound pairs and
unpaired fermions with ferromagnetic order.
The external field-magnetization relation
\begin{eqnarray}
\frac{H}{2}& \approx &\frac{\hbar^2n^2}{2m}\left\{\frac{\gamma^2}{4}
+4\pi^2(m^z)^2\left(1+\frac{4(1-2m^z)}{|\gamma|}-\frac{8m^z}{3|\gamma|}\right)\right.\nonumber\\
& &
\left.-\frac{\pi^2}{16}\left(1-2m^z\right)^2\left(1+\frac{8m^z}{|\gamma|}\right)
\right\},\label{H}
\end{eqnarray}
follows by solving equations (\ref{Mu-b}) and (\ref{Mu-u}).
It indicates the energy transfer relation among the kinetic energy
variation $\Delta E_k=\mu^{\rm u}-\mu^{\rm b}$, the binding energy
$\epsilon_{\rm b}$ and the Zeeman energy $\mu_{B}H$: 
\begin{eqnarray}
\Delta E_{k} +\epsilon_{\rm b} =\mu_{B}H \label{H-E-relat}
\end{eqnarray}
 which qualitatively agrees with the relation 
identified in experiment \cite{Fermi-1D1}. In the above equation we
take the Bohr magneton  $\mu_{B}=1$ and $m^z=M^z/n$.
\begin{figure}[t]
{{\includegraphics [width=1.0\linewidth]{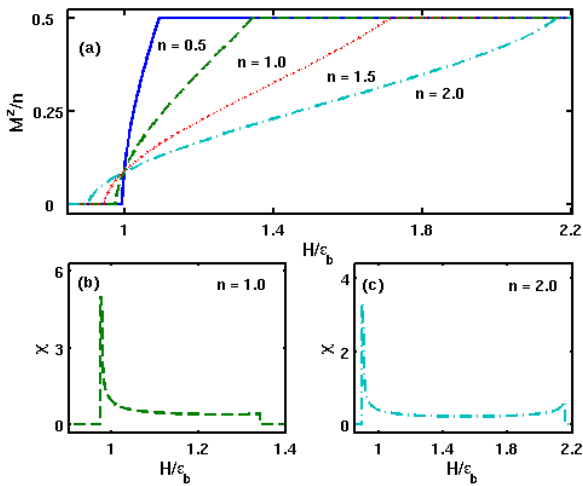}}}
\caption{Magnetization $M^z$ and susceptibility $\chi$ vs the
external field $H$ in the units $2m=\hbar=1$, according to Eq.~(\ref{H}).}\label{fig:Mz}
\end{figure}

Figure \ref{fig:Mz}(a) shows the magnetization and the
susceptibility for $ |c|=10$ and $n=0.5,1,1.5,2$. The
magnetization gradually increases from $M^{z} = 0$ 
to $n/2$ as the field increases from $H_{c1}$ to $H_{c2}$.
It is important to note that the points of intersection
at $H=\epsilon_{\rm b}$ indicate where the Fermi surface of
unpaired fermions exceeds the one for the bound pairs.  This point
separates the mismatched pairing phase into different breached pairing
phases \cite{Sama,MF-FFLO}.
The susceptibility shows discontinuities at the critical points, with 
$\chi=0$ for $H<H_{c1}$ and $H>H_{c2}$.
However, $\chi$ is finite and quickly decreases in the vicinity of $H_{c1}$.
For larger densities (Figure \ref{fig:Mz}(c) illustrates the case $n=2$),
$\chi$ slowly increases as $H \to H_{c2}$.
We note that the coexistence of pairing and magnetization in the 1D
attractive Fermi gas is similar to the Shubnikov phase of
superconductivity and magnetization in type II superconductors \cite{Super-II}.

\begin{figure}[t]
{{\includegraphics [width=0.98\linewidth]{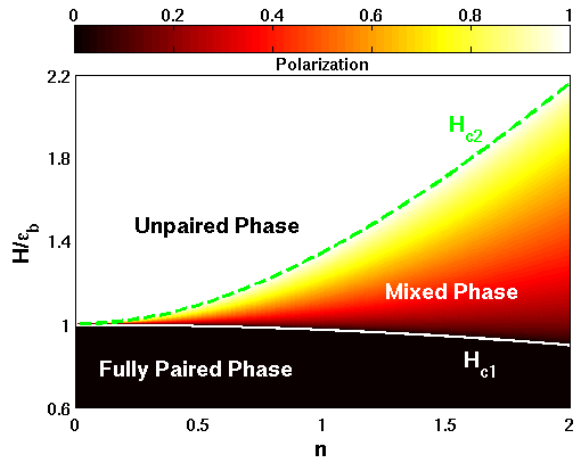}}}
\caption{Phase diagram for homogeneous systems with $|c| = 10$ 
according to Eqs.~(\ref{Hc1}), (\ref{Hc2}) and (\ref{H}).}\label{fig:PD}
\end{figure}

Figure \ref{fig:PD} shows the phase diagram in the $n-H$ plane for the particular value $|c|=10$.
As $n \to 0$, the two critical fields approach the binding energy $\epsilon_{\rm b}$.
The two critical fields have opposite monotonicity: $H_{c1}$ decreases with increasing $n$
whereas $H_{c2}$ increases with $n$.
For the 1D Fermi gas in a harmonic trapping potential
\cite{Orso,Hu}, the density is position-dependent and decreases
away from the trapping centre.
Thus, for sufficiently large centre-density, the system has subtle
segments; the mixed phase lies in the centre and the fully paired
phase (or the fully unpaired phase) sits in the two outer wings
for $H < \epsilon_{\rm b}$ (or $H > \epsilon_{\rm b}$).
Nevertheless, for sufficiently low centre-density, the cloud is
either a fully paired phase or a fully unpaired phase for 
$H < \epsilon_{\rm b}$ or $H > \epsilon_{\rm b}$, respectively.

For the 1D Fermi gas, the local pair correlation is defined by
\begin{equation}
g^{(1)}_{\rm p} = \left \langle \phi
_{\downarrow}^{\dagger}(0) \phi _{\uparrow}^{\dagger}(0) \phi
_{\uparrow}^{}(0) \phi _{\downarrow}^{}(0) \right \rangle \approx
\frac{1}{2}\frac{dE}{dc}.
\end{equation}
For weakly attractive interaction, 
\begin{equation}
g^{(1)}_{\rm p}\approx n^2 (1-P^2)/4
\end{equation}
indicating a two-component free Fermi gas phase.
For strongly attractive interaction the local pair correlation is given by
\begin{equation}
 \frac{g^{(1)}_{\rm p}}{n^2} \approx
\frac{(1-P)}{4}\left[|\gamma|+\frac{\pi^2(1-P)^2(1+3P)}{24\gamma^2}+\frac{8\pi^2P^3}{3\gamma^2}\right].
\end{equation}
This has maximum and minimum values corresponding to the fully
paired phase for $H<H_{c1}$ and the fully unpaired phase for $H>H_{c2}$, respectively.
Depairing weakens the pair correlation in the region  $H_{c1}<H<H_{c2}$.
The phase transitions in the vicinities of the critical points are of second order.

\section{Exclusion Statistics}
\label{GES}

Strong thermal fluctuations can destroy the magnetically ordered phases,  
delineated by the two critical fields, above a critical temperature $T_c$, 
which in principle can be also calculated from the TBA equations.  
On the other hand, at temperatures much lower
than the degeneracy temperature $T_d:=\frac{\hbar^2}{2m}n^2\ll
\epsilon_{\rm b}$, the bound
pairs are stable against weak thermal fluctuations. 
However, the individual pair wavefunctions do not overlap coherently, i.e., the
existence of bound pairs does not lead to long range order at finite temperatures.
This can be seen from the finite temperature TBA equations
(\ref{TBA-Full})  in which the unpaired band is empty due to a large
negative chemical potential. This behaviour is similar to that of 3D
attractive fermions \cite{Chin}.
In 1D the dynamical interaction and
the statistical interaction in the pairing scattering process are
inextricably related \cite{Wu,Haldane,Wilczek,BGO}.
This means that one bound pair excitation may cause a fractional number of holes
below the Fermi surface due to the collective signature.  
This is the key point in understanding GES for 1D interacting many-body
systems. We shall show that the bound pairs can be
viewed as ideal particles obeying GES. 
GES has recently been applied to the 3D unitary Fermi gas \cite{Muthry}.

In the absence of the magnetic field and at low temperatures, the
unpaired dressed energy is  positive due to a large negative
chemical potential. 
It follows that the BA and TBA equations can be written as
\begin{eqnarray}
\sigma(k )+\sigma^h(k )
&=&\frac{1}{\pi}-\int_{-\infty}^{\infty}a_2(k-\Lambda)\sigma(\Lambda)d\Lambda,\label{BA-d}\\
\epsilon^{\rm b}(k )&=&2(k^2-\mu -\frac14{c^2})\nonumber\\
&& \,+\, Ta_2*\ln\left(1+{\mathrm e}^{-\frac{\epsilon(k)}{K_BT}}\right).\label{TBA-F2}
\end{eqnarray}

After neglecting  exponentially small terms in (\ref{P-b}),  the pressure for the bound
pairs is given by
\begin{eqnarray}
p^{\rm b}\approx
\frac{2}{\sqrt{\frac{2\pi^2\hbar^2}{2m}}}\int_0^{\infty}\frac{\sqrt{\epsilon}d\epsilon
}{1+\mathrm{e}^{\frac{\epsilon-2A(T)}{K_BT}}}\label{Pressure-2}
\end{eqnarray}
with the function $A(T):= \hbar ^2B^2/(2m)=(\mu+\frac14{c^2}-p^{\rm b}/2c)$.  
Furthermore, using Sommerfeld
expansion and iterating the pressure $p^{\rm b}$ with equation
(\ref{TBA-F2}), we obtain the cut-off energy $A(\tau)$ in the form
\begin{equation}
A(\tau) \approx A_0\left[1+\frac{16\tau^2}{3\pi^2} \left(1-\frac{2}{|\gamma|} \right)
+\frac{1024\tau^4}{9\pi^4} \left(1-\frac{4}{|\gamma|} \right)\right], \label{TBA-mu}
\end{equation}
where
\begin{equation}
A_0=\frac{\hbar^2}{2m}\frac{n^2\pi^2}{16} \left(1+\frac{1}{|\gamma|} \right). 
\end{equation}
Here $\tau=K_BT/T_{d}$ is  the degenerate temperature.  
The pair distribution function $n(\epsilon):= \pi \sigma(\epsilon)$ is given by
\begin{equation}
n(\epsilon)=\frac{1}{\alpha(1-\mathrm{e}^{(\epsilon-2A(k)/K_BT})}, \label{TBA-n}
\end{equation}
where $\alpha =1+1/|2\gamma|$.

The chemical potential follows as 
\begin{eqnarray}
\mu &\approx& \mu_0\left[1+\frac{16\tau^2}{3\pi^2} \left(1-\frac{4}{3|\gamma|}\right)+
\frac{1024\tau^4}{9\pi^4} \left(1-\frac{56}{15|\gamma|} \right) \right]  \nonumber\\
&&\,-\frac12{\epsilon_B}.
\end{eqnarray}
Here 
\begin{equation}
\mu_0 \approx \frac{\hbar^2}{2m}\frac{n^2\pi^2}{16}\left(1+\frac{4}{3|\gamma|}\right), 
\end{equation}
which is consistent with (\ref{Mu-b}).  
The total energy per unit length and the free
energy per unit length in the strong coupling regime are
\begin{eqnarray}
E&\approx &E_0\left[1+\frac{16\tau^2}{\pi^2} \left(1-\frac{2}{|\gamma|} \right)+
\frac{1024\tau^4}{5\pi^4} \left(1-\frac{4}{|\gamma|} \right)\right] \nonumber\\
&& \,-\,\frac12 {n \epsilon_{\rm b}}, \nonumber\\ 
F&\approx & E_0\left[1-\frac{16\tau^2}{\pi^2} \left(1-\frac{2}{|\gamma|} \right)
-\frac{1024\tau^4}{15\pi^4} \left(1-\frac{4}{|\gamma|}\right)\right] \nonumber\\
&& \,-\, \frac12 {n\epsilon_{\rm b}},\label{TBA-EF}
\end{eqnarray}
respectively. 
Here 
\begin{equation}
E_0=\frac{\hbar^2n^3}{2m} \frac{\pi^2}{48}\left(1+\frac{1}{|\gamma|}\right)
\end{equation}
is consistent with (\ref{e0}) obtained from (\ref{TBA-F}). 

We see that in the strongly
attractive regime and in the absence of the magnetic field the bound pairs
behave like hard-core bosons at low temperatures and have massless
excitations, i.e.
\begin{equation}
F(T)=F(0)-\frac{\pi C(K_BT)^2}{6\hbar v_c} +O(T^4). \label{F-T}
\end{equation}
Here the central charge $C=1$ and $v_c$ is given by (\ref{vc}).
The specific heat is given by 
\begin{eqnarray}
c_v=\frac{nK_B\tau}{3(1+\frac{1}{|\gamma|})}
\left(1 + \frac{128\tau^2}{5\pi^2}\left(1-\frac{2}{|\gamma|}\right)\right).\label{F-T}
\end{eqnarray}

On the other hand, the statistical signature of the fully paired state
can be described by GES \cite{Haldane,Wu}.
In this formalism the pair distribution function is given by 
\begin{equation}
n(\epsilon)\approx({\alpha+w(\epsilon)})^{-1}\label{GES-n}
\end{equation}
where $w(\epsilon)$ satisfies the GES relation 
\begin{equation}
w^{\alpha}(\epsilon)\left(1+w(\epsilon)\right)^{1-\alpha} = {\rm e}^{\frac{\epsilon-2A(T)}{K_BT}}.  
\end{equation}
Here $\epsilon$ denotes the
energy of pairs.  
For the strongly attractive Fermi gas, we find the GES parameter
\begin{equation}
\alpha \approx 1+1/|2\gamma|.
\end{equation}

\begin{center}
\begin{figure}[ht]
\includegraphics[width=0.90\linewidth]{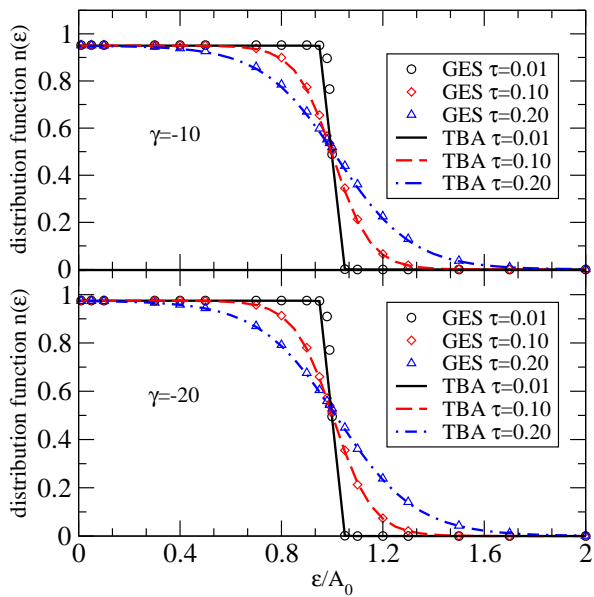}
\caption{Comparison between the most probable distribution profiles
  $n(\epsilon)$ for the values $\gamma=-10$ and $-20$ at different values
  of the degeneracy temperature $\tau=K_BT/T_d$.  At zero temperature
  $n(\epsilon)=1/\alpha $ leads to a Fermi surface at $\epsilon =2A_0$.
  Fermi-Dirac statistics with GES parameter $\alpha=1$ appear for
  $\gamma \to \infty$. Attractive interaction thus results in a more
  exclusive state than for pure Fermi-Dirac statistics.  The solid and dashed lines
  are obtained from the TBA distribution function (\ref{TBA-n}).  The symbols show
  the most probable distribution evaluated from the GES result
  (\ref{GES-n}).  The results from both approaches are seen to coincide well. }
\label{fig:n}
\vspace{0.5cm}
\end{figure}
\end{center}

\begin{center}
\begin{figure}[ht]
\vskip 5mm
\includegraphics[width=1.0\linewidth]{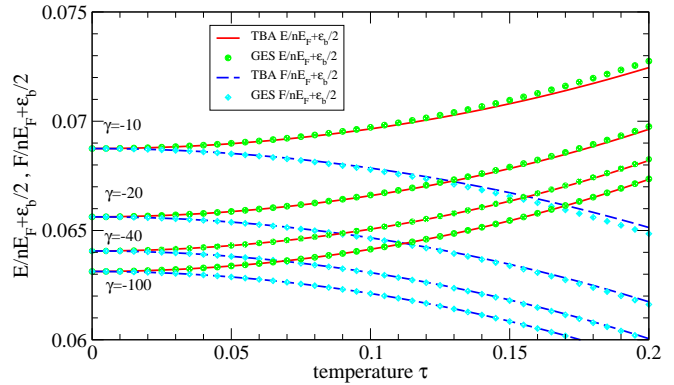}
\caption{Free energy per unit length $F+\epsilon_{\rm b}/2$ and total
  energy per unit length $E+\epsilon_{\rm b}/2$ in units of 
  $nE_{\rm F}=\frac{\hbar^2}{2m}\pi^2n^3/3$ {\it vs} the degenerate temperature
  $\tau=K_BT/T_d$ for different coupling strengths $\gamma =-10, \, -20,\, -40,\, -100$.  
  The solid and dashed lines are evaluated from the TBA results
  (\ref{TBA-EF}).  The symbols follow from the GES
  results (\ref{GES-EF}).  For $\gamma =-10$, a small discrepancy
  between the TBA and GES results is observed due to the approximation
  condition $\gamma \gg 1$. In the strong coupling
  regime, the universality (\ref{F-T}) of low temperature behaviour
 suggests that strongly attractive fermions can be viewed 
  as ideal particles obeying nonmutual GES with $\alpha \approx
  1+1/|2\gamma|$ for temperatures $K_BT\ll \epsilon_{\rm b}$.}
\label{fig:EF}
\end{figure}
\end{center}

Now following Isakov {\em et al.} \cite{Isakov2}, at low temperatures,
i.e., for $K_BT < T_d$, we find the cut-off energy 
\begin{eqnarray}
A(T)\approx A_0\left[1+\frac{16\tau^2}{3\pi^2\alpha^3}+O(\tau^4) \right],\nonumber
\end{eqnarray}
which agrees with (\ref{TBA-mu}) to leading order and next leading
order in the strong coupling regime 
(for higher order terms, the reader is referred to Ref.~\cite{BGO}).
Figure \ref{fig:n} shows the close agreement between the TBA
distribution function (\ref{TBA-n}) and the GES most probable
distribution of fermion pairs (\ref{GES-n}) for different values of
interacting strength at low temperatures.  
We see clearly that the dynamical interaction $\gamma$ continuously varies the GES, 
with the most probable distribution of fermion pairs approaching that of 
hard-core bosonic molecues with an effective statistics parameter
$\alpha=1$ as the interaction increases.  
In this sense the dynamical attractive interaction makes the fermions more exclusive.  
In the GES formalism the total energy per unit length and the free energy per unit length 
follow as
\begin{eqnarray}
E&\approx &E_0\left[1+\frac{16\tau^2}{\pi^2\alpha^3}+O(\tau^4)\right]
-\frac12\, {n \epsilon_{\rm b}},\nonumber\\ 
F&\approx &E_0\left[1-\frac{16\tau^2}{\pi^2\alpha^3}+O(\tau^4)\right]
-\frac12\, {n \epsilon_{\rm b}},\label{GES-EF}
\end{eqnarray}
which again agree well with the TBA results (\ref{TBA-EF})  for strong coupling, see Figure \ref{fig:EF}.

\section{Conclusion}
\label{conclusion}

In conclusion, we have studied pairing and quantum phase transitions in the 
strongly attractive 1D Fermi gas with an external magnetic-like field. 
Analytic results have been obtained for the critical fields $H_{c1}$ and $H_{c2}$, 
magnetization, critical behaviour and local pair correlation.
The pairing induced by an interior gap in the system differs from
conventional BCS pairing and gapped spin liquids.
The smooth pair breaking phase transitions seen in the attractive
Fermi gas are reminiscent of the superconductivity breaking phase transitions
in type II superconductors \cite{Super-II}.
At low temperatures we predict that the hard-core bound pairs of
fermionic atoms obey GES. 
The thermodynamics of the hard-core pairs obey universal
temperature dependent scaling.  


We emphasize here that in the presence of an external magnetic field, 
pair breaking in the 1D two-component strongly attractive Fermi
gas sheds light in understanding the pairing signature of the 3D
strongly interacting Fermi gas of ultracold atoms in which superfluid
and normal phases can coexist.  
Although there is no long range order in 1D quantum many-body physics, the
mismatched Fermi surfaces do not prevent pairing. 
This pairing signature has
also been observed in the 3D two-component atomic gas with high spin
population imbalances \cite{Schunck}.  
In addition, for the 1D Fermi gas, the magnetic field triggers spin imbalances when 
the external field is greater than the first critical field $H_{c1}$ (\ref{Hc1}) 
and less than the second critical field $H_{c2}$ (\ref{Hc2}). 
The energy transfer relation (\ref{H-E-relat}) found for this model is also consistent
with experimental observations in the 1D Fermi gas \cite{Fermi-1D1}. 
The phase diagram presented in Figure \ref{fig:PD} clearly shows the
phase separation and the pairing signature with changing magnetic field.
It may be possible to experimentally test our theoretical predictions for the quantum phase
transitions and critical fields in the 1D two-component strongly interacting Fermi gas via 
the experimental advances in trapping ultracold atoms.

This work has been supported by the Australian and German Research
Councils. The authors thank Prof. M. Takahashi for helpful discussions and
C.L. thanks Prof. Yu.S. Kivshar for support.

\clearpage

\end{document}